\documentclass[a4paper,11pt]{article}
\pdfoutput=1 

\usepackage{jcappub} 

\usepackage[T1]{fontenc} 
\usepackage{color}
\definecolor{Red}{rgb}{0.65,0.08,0.05}
\definecolor{Blue}{rgb}{0.05,0.08,0.65}

\newcommand{\ze}{{z}}

\newcommand{\dd}{{\rm d}}
\newcommand{\ii}{{\rm i}}

\newcommand{\vx}{\textbf{x}}
\newcommand{\vd}{\textbf{d}}

\newcommand{\vq}{\textbf{q}}
\newcommand{\vk}{\textbf{k}}

\newcommand{\mH}{{\cal H}}

\newcommand{\mS}{{\cal S}}

\newcommand{\Dirac}{{\delta_{\rm Dirac}}}

\newcommand{\adiab}{{\rm adiab}}

\newcommand{\cdm}{{\rm cdm}}
\newcommand{\vPi}{{\tau}}

\begin{document}
\title{On the importance of nonlinear couplings in large-scale neutrino streams}

\author[a,b]{H\'{e}l\`{e}ne Dupuy,}
\author[a,b]{Francis Bernardeau}

\affiliation[a]{Institut d'Astrophysique de Paris, UMR-7095 du CNRS, Universit\'e Pierre et Marie Curie, 98 bis bd Arago, 75014 Paris, France}
\affiliation[b]{Institut de Physique Th\'eorique, CEA, IPhT, F-91191 Gif-sur-Yvette,\\
CNRS, URA 2306, F-91191 Gif-sur-Yvette, France}

\emailAdd{helene.dupuy@cea.fr}      
\emailAdd{francis.bernardeau@iap.fr}

\abstract{We propose a procedure to evaluate the impact of nonlinear couplings
on the evolution of massive neutrino streams in the context of large-scale structure growth. 
Such streams can be described by general nonlinear conservation equations, derived from a multiple-flow perspective, which generalize the conservation equations of non-relativistic pressureless fluids. 
The relevance of the nonlinear couplings is quantified with the help of the eikonal approximation applied to the subhorizon limit of this system. It highlights the role played by the relative displacements of different cosmic streams
and it specifies, for each flow, the spatial scales at which the growth of structure is affected by nonlinear couplings. We found that, at redshift zero, such couplings can be significant for wavenumbers as small as $k=0.2\,h$/Mpc for most of the neutrino streams.}

\date{\today}
\maketitle

\bigskip


\section{Introduction}

Statistical properties of the large-scale structure of the universe have long been proposed 
as an efficient instrument to constrain cosmological parameters. In this context,
a careful account of the role played by massive neutrinos is crucial. So far, it has often been overlooked in the nonlinear or quasilinear regime because of the technical complexity specific to the study of massive neutrinos (see recent attempts in \cite{2009JCAP...06..017L,2010PhRvD..81l3516S,2014arXiv1408.2995B}). 
This is all the more unfortunate that cosmological observations can fruitfully improve our knowledge of those particles. 
The signature of neutrino masses on cosmological observables is indeed expected to be significant enough for those masses to be constrained observationally
\cite{2011PhRvD..83d3529S,2012PhRvD..85h1101R,2013JCAP...01..026A,2011JCAP...03..030C,2011arXiv1110.3193L,2009A&A...500..657T}. 

In the linear regime, the effect of neutrinos is now well understood (see refs. \cite{1994ApJ...429...22M,Ma:1995ey,Lesgourgues2006}). The need for nonlinear corrections in their equations of motion has been raised because the cosmological 
observations that
are the most sensitive to neutrinos masses, i.e. for wavenumbers in the $0.1$-$0.2\ h$/Mpc range, precisely correspond to the
mildly nonlinear regime. To deal with this issue, several strategies can be adopted. Until recently, in analytic works, the neutrino fluids had always been treated in the linear regime, nonlinear couplings being introduced in the dark matter
description only. This nonlinear treatment can be implemented with the help of the Renormalization Group time-flow approach \cite{2008JCAP...10..036P,2009JCAP...06..017L}.
Improvements upon such schemes have been proposed in \cite{2014arXiv1408.2995B}, which consists in a hybrid approach that matches the full Boltzmann hierarchy to an effective two-fluid description at an intermediate redshift.  \cite{2014arXiv1412.2764F}  is, for its part, a systematic perturbative expansion of the Vlasov equation in which high-order corrections to the neutrino density contrast are computed without the explicit need to track the perturbed neutrino momentum distribution.

Ideally, however, the fully nonlinear evolution of the neutrino fluid should be depicted. A natural way to do so would be
to take inspiration of the standard linear description, which relies on the Boltzmann equation, and extend the harmonic decomposition of the phase-space distribution function to the nonlinear regime. This has been done in \cite{vanderijtphd} but this method turned out to be particularly difficult to handle. In  \cite{2014JCAP...01..030D,2014arXiv1411.0428D}, we proposed to describe massive neutrinos as a superposition of single-flow fluids, the equations of motion of each of them being written in the nonlinear regime.

In this paper, we are interested in exploiting those theoretical developments in order to identify the scales at which nonlinear couplings in the neutrino fluids are expected to play a significant role. In order to do so, we apply the eikonal approximation to the nonlinear equations of motion. Note that this approximation had already been exploited in the literature to develop a Perturbation Theory approach for cold dark matter  (\cite{2012PhRvD..85f3509B}) and had proved to be able to capture the leading coupling effects.

The article is structured as follows. In section \ref{Eom}, we recall the form of the nonlinear equations of motion describing the time evolution of non-interacting fluids, relativistic or not, when using a multi-fluid approach. Section \ref{eikonal approximation} explains in detail how the eikonal approximation can be implemented in those equations and emphasizes the key role of relative displacement fields. Finally, in section \ref{power spectra}, power spectra of the relative displacements between neutrino fluids and the cold dark matter component are  presented. The impact on the growth of large-scale structure is then discussed in a quantitative way.

\section{Nonlinear equations of motion (multi-fluid description)}\label{Eom}
Following \cite {2014JCAP...01..030D,2014arXiv1411.0428D}, it is now clear that any non-interacting relativistic fluid can be divided into several flows, each of them evolving then independently until first shell-crossings. In cosmology, this approach obviously applies to massive neutrinos since they are free-streaming. In this framework, each flow can be defined as the collection of all the particles (with a mass $m$) having the same comoving momentum. They are entirely characterized by two coupled fields, namely the comoving number density $n_{c}$ and momentum $P_{\mu}$. Those fields obey the following general equations,
\begin{equation}
\frac{\partial}{\partial\eta}n_{c}+\frac{\partial}{\partial x^{i}}
\left(\frac{P^{i}}{P^{0}}n_{c}\right)=0,\label{ncevol}
\end{equation}
with $P^{\mu}(\eta,x^{i})=g^{\mu\nu}P_{\nu}(\eta,x^{i})$  and $P^\mu P_\mu=-m^2$, $g_{\mu\nu}$  being the metric, and
\begin{equation}\label{GeneralPiEvol1}
P^\nu P_{\mu,\nu}=\dfrac{1}{2} P^\sigma P^\nu g_{\sigma \nu,\mu}.
\end{equation}
These equations directly ensue from the matter and momentum conservation equations, applied to each flow.
At this stage,  no perturbative expansion of the metric is involved.
The properties of the whole fluid are then inferred by examining an appropriate number of such flows, each of them being labeled
by the initial value of its field $P_i$, denoted $\vPi_{i}$  (found to be constant at zeroth order in Perturbation Theory). The initial number density
of particles in each flow is constrained by the choice of  initial conditions. For instance, the case of adiabatic initial conditions
is described in detail in  \cite{2014JCAP...01..030D}.

We are interested here in equations i) involving linearized metric perturbations (but non-linearized fields) and ii) 
rid of the coupling terms that are subdominant at subhorizon scales. They can be written in terms of the number density contrast
\begin{equation}
\delta_{\vPi_{i}}(\eta,x^{i})=\dfrac{n_c(\eta,x^{i};\vPi_{i})}{n_c^{(0)}(\vPi_{i})}-1
\end{equation}
and of the velocity divergence field (in units of $- \mathcal{H}$, $\mathcal{H}$ being the conformal Hubble constant)
\begin{equation}
\theta_{\vPi_{i}}(\eta,x^{i})=-\frac{\partial_i P_{i}(\eta,x^{i};\vPi_{i})}{m a\mH}
\end{equation}
since the field $P_{i}$ has found to be potential in this regime\footnote{This property, rigorously demonstrated in \cite{2014arXiv1411.0428D}, generalizes that of non-relativistic species.}.
For a generic perturbed Friedmann-Lema\^{i}tre metric\footnote {Units are chosen so that the speed of light in vacuum is equal to unity.} , whose time variable is the conformal time $\eta$,
\begin{equation}
\dd{s^2}=a^2(\eta)\left[-(1+2A)\dd{\eta^2}+2B_i\dd{x^i}\dd{\eta}+(\delta_{ij}+h_{ij})\dd{x^i}\dd{x^j}\right],
\end{equation}
one has in Fourier space for the mode characterized by the wave vector $\vk$ (\cite{2014arXiv1411.0428D})
\begin{eqnarray}
\left(a\partial_a-\ii\dfrac{\mu k \tau }{\mH \tau_0}\right)
\delta_{\vPi_{i}}(\vk)
+\frac{m a}{\vPi_{0}}\left(1-\dfrac{\mu^{2}\tau^2}{\tau_0^2}\right)\theta_{\vPi_{i}}(\vk)
&=&\nonumber\\
&&\hspace{-3cm}-\dfrac{m a}{\tau_0}\int{\dd^3\vk_{1} \dd^3\vk_{2}}\alpha_{R}(\vk_{1},\vk_{2};\vPi_{i})\delta_{\vPi_{i}}(\vk_{1})\theta_{\vPi_{i}}(\vk_{2}),
\label{Fouriersub1}\\
\left(1+a \dfrac{\partial_a \mH}{\mH}+a \partial_a-\ii \dfrac{\mu k \tau}{\mH\tau_0 }\right)\theta_{\vPi_{i}}(\vk)-\dfrac{k^2}{m a\mH^2}\mS_{\vPi_{i}}(\vk)&=&
\nonumber\\
&&\hspace{-3cm}-\dfrac{m a}{\tau_0}\int{\dd^3\vk_{1} \dd^3\vk_{2}}
\beta_{R}(\vk_{1},\vk_{2};\vPi_{i})\theta_ {\vPi_{i}}(\vk_{1})\theta_ {\vPi_{i}}(\vk_{2}).
\label{Fouriersub2}
\end{eqnarray}
We have introduced here
\begin{equation}
\mu=\frac{k_{i}\tau_{i}}{k\tau},\tau^{2}=\tau_{i}\tau_i
 \text{ and }
\tau_{0}=-\sqrt{m^{2} a^2+\vPi_{i}^{2}}.
\end{equation}
Besides, $\mS_{\vPi_{i}}(\vk)$ is a source term given by
\begin{equation}
\mS_{\vPi_{i}}(\vk)=\tau_0 A(\vk)+\vec{\tau}\cdot\vec{B}(\vk)-\frac{1}{2}\frac{\vPi_{i}\vPi_{j}}{\vPi_{0}}h_{ij}(\vk).
\end{equation}
These equations contain also the generalized kernel functions, adapted to relativistic flows,
\begin{align}\label{alphaR}
\alpha_{R}(\vk_{1},\vk_{2};\vPi)=\Dirac(\vk-\vk_{1}-\vk_{2})\dfrac{(\vk_{1}+\vk_{2})}{k_2^2}\cdot\left[\vk_{2}-\vec{\tau}\dfrac{\vk_{2}\cdot \vec{\tau}}{\tau_0^2}\right],
\end{align}
\begin{align}\label{betaR}
\beta_{R}(\vk_{1},\vk_{2};\vPi)=\Dirac(\vk-\vk_{1}-\vk_{2})\dfrac{\left(\vk_{1}+\vk_{2}\right)^{2}}{2k_1^2 k_2^2}\left[\vk_{1} \cdot \vk_{2}-\dfrac{\vk_{1}\cdot \vec{\tau}\vk_{2}\cdot \vec{\tau}}{\tau_0^2}\right].
\end{align}
As mentioned in \cite{2014arXiv1411.0428D}, they are extensions of the kernel functions found for pressureless fluids of
non-relativistic species (see \cite{2002PhR...367....1B} for details in this context.).

In any practical implementation, it is necessary to consider a collection 
of $N$ streams. In that case, the general equation of motion can  conveniently be written  in terms of the time-dependent $2N$-uplet,
\begin{equation}
\Psi_{a}(\vk)=(\delta_{\vPi_1}(\vk),\theta_{\vPi_{1}}(\vk),\dots,\delta_{\vPi_{N}}(\vk),\theta_{\vPi_{N}}(\vk))^{T}.
\end{equation}
Before shell-crossing, it can incorporate all the relevant species (neutrinos, dark matter, baryons) as long as they interact only via gravitation.
In this context, the equations (\ref{Fouriersub1}) and (\ref{Fouriersub2}) of all the flows 
can formally be recast in the form\footnote{The Einstein notation for the summation over repeated indices is adopted and, in the right hand side of this equation, it is
assumed that the wave modes are integrated over.}
\begin{equation}
\partial_{\ze}\Psi_{a}(\vk)+\Omega_{a}^{\ b}\,\Psi_{b}(\vk)=\gamma_{a}^{\ bc}(\vk_{1},\vk_{2})\Psi_{b}(\vk_{1})\Psi_{c}(\vk_{2}),
\label{FullEoM}
\end{equation}
where 
${\partial}/{\partial \ze}\equiv {a\partial}/{\partial a}$ and
the indices $a$ and  $b$ run from 1 to $2N$.
The matrix elements $\Omega_a^{\ b}$ gather the linear couplings. They contain in particular the way in which the source terms $\mS_{\vPi_{i}}(\vk)$  can be re-expressed as a function of the $2N$-uplet elements. The left hand side of this equation is nothing but the
linear field evolution. The right hand side contains the coupling terms. More precisely,
the \emph{symmetrized vertex} matrix $\gamma_{a}^{\ bc}(\vk_{1},\vk_{2})$ describes the nonlinear 
interactions between the Fourier modes. It is given by
\begin{equation}\label{vertex1}
\gamma_{2p-1}^{\ 2p-1\,2p}(\vk_1,\vk_2)=-\frac{ma}{2{(\vPi_p)}_{0}}
\alpha_{R}(\vk_{1},\vk_{2},\vPi_{p}),\\
\end{equation}
\begin{equation}
\gamma_{2p}^{\ 2p\,2p}(\vk_1,\vk_2)=
-\frac{ma}{{(\vPi_p)}_{0}}
\beta_{R}(\vk_{1},\vk_{2},\vPi_{p}),
\label{vertex2}
\end{equation}
with $\gamma_{a}^{\ bc}(\vk_1,\vk_2)=\gamma_{a}^{\ cb}(\vk_2,\vk_1)$ and $\gamma_{a}^{\ bc}=0$ 
otherwise. 

\section{The eikonal approximation}\label{eikonal approximation}

The eikonal approximation\footnote{In this context, the term refers to diagram resummations performed in quantum electrodynamic field
equations, \cite{Abarbanel:1969ek}.}, developed in 
\cite{2012PhRvD..85f3509B},
is based on the observation that the amplitudes of the kernel functions describing mode couplings, $\alpha_{R}$  and $\beta_{R}$, 
significantly depend on  the ratio between the wave numbers at play. This has been observed for non-relativistic fluids and we show here that it is the case for relativistic fluids too. This property leads to the idea that the 
right hand side of eq. (\ref{FullEoM})  can be split into two integration domains. One is called the hard domain and encompasses modes whose wavelengths are of the same order (and for which the coupling functions are always finite).
The other one is referred to as the soft domain. It is made of  modes 
of very different wavelengths for which the coupling functions, of the order of the wavelength ratio, are large. 

The main idea is that there are regimes in which the dominant coupling structure is in the soft domain. This is the case for instance in early-time  fluids containing baryons and cold dark matter. At the time of recombination, the baryon velocity drops steeply whereas the velocity of cold dark matter is not affected by decoupling. It means that, at intermediate scales (i.e. between the Silk damping length and the sound horizon), the relative velocity $v_{\rm rel}$ between the baryon flow and the one of cold dark matter is substantial. Because of this, for scales at which $k>aH/\langle v_{\rm rel}^2 \rangle^{1/2}$ (at decoupling), the wavelength of gravitational potential wells is too small for baryons to fall in them before  being pushed towards another direction. Eventually, this phenomenon induces a damping of the matter power spectrum. The relative motion between cold dark matter and baryons and its effects on the matter power spectrum have been highlighted in \cite{2010PhRvD..82h3520T,2013PhRvD..87d3530B}. Such studies illustrate the relevance of couplings between large scale modes and small scale modes in the framework of structure formation. For instance, in the case of mixtures of baryons and cold dark matter, the typical coherence length of the relative velocity field is of the order of few Mpc, which is much larger than the scales at which basic baryonic objects start to form under gravitational clustering (of the order of 10 kpc). The same formalism can be used to obtain the large $k$ behavior of the
propagators in case of a single pressureless fluid, reproducing the results obtained in \cite{2006PhRvD..73f3520C,2010PhRvD..82h3507B}.
In the same spirit, we propose here to investigate the impact of the relative motion between given neutrino streams and the cold dark matter fluid with the help the eikonal approximation. It will allow us to infer the amplitude of neutrino coupling effects. 

For convenience, let us assume that the soft domain is obtained for $k_{2}\ll k_{1}$ in eq. (\ref{FullEoM}). We have then $\vk=\vk_{1}$ and the contribution corresponding to the soft domain can be viewed as a mere corrective term
in the \emph{linear} equation describing the evolution of the mode $\vk$. In other words, eq. (\ref{FullEoM}) can be rewritten 
\begin{eqnarray}\label{EikEoM}
&&\frac{\partial}{\partial \ze} \Psi_a(\ze,\vk) + \Omega_{a}^{\ b}(\ze,\vk) \Psi_b(\ze,\vk) -\Xi_{a}^{b}(\ze,\vk)\Psi_{b}(\ze,\vk)\nonumber\\
&&\hspace{5cm}=\left[\gamma_{a}^{\ bc}(\vk_1,\vk_2) \ \Psi_b(\vk_1,\ze) \ \Psi_c(\vk_2,\ze)\right]_{\mH},
\end{eqnarray}
with
\begin{equation}
\Xi_{a}^{\ b}(\vk,\ze)\equiv  2 \int_{{\mS}}\dd^{3}\vq \; 
\,^{\rm eik.}\gamma_{a}^{\ bc}(\vk,\vq)
\Psi_{c}(\vq,\ze)  \;. \label{fb:Xidef}
\end{equation}
The soft momenta $q$ (i.e. $q\ll k$) at play in eq. (\ref{fb:Xidef}) are integrated over so that 
$\Xi_{a}^{\ b}(\ze,\vk)$ is independent on $\Psi_{a}(\ze,k)$. It is a mere time and scale dependent matrix. 
The fact that the integration domain is restricted 
to the soft wave numbers in eq. (\ref{fb:Xidef}) is the key element. Conversely, in the right-hand side of eq. (\ref{EikEoM}), the implicit convolution product excludes the soft domain (i.e. all the modes  concerned have comparable wavelengths).  When the contribution of the hard domain is negligible, \emph{eq. (\ref{EikEoM}) can then be viewed as the equation of motion of the mode $\mathbf{k}$ evolving in a medium perturbed by large-scale modes.}
It therefore encodes the way in which long-wave modes alter the growth of structure. 
Once $\Xi_{a}^{b}(\ze,\vk)$ is given, eq. (\ref{EikEoM}) can be solved as a linear equation. This is precisely the eikonal approximation
of the global equation of motion.

In practice, applying the eikonal approximation to $\Xi_{a}^{\ b}(\vk,\ze)$ means that the vertex values that appear in this quantity have to be computed assuming $k_{2}\ll k_{1}$. In this framework, one deduces from eqs. (\ref{alphaR}), (\ref{betaR}), (\ref{vertex1}) and (\ref{vertex2}) that the eikonal limit of the vertex elements is
\begin{equation}\label{gamma1}
\,^{\rm eik.}\gamma_{2p}^{bc}(\vk,\vk_{2})=-\delta_{2p}^{b}\delta_{2p}^{c}\ \frac{m a}{2k_{2}^{2}{(\vPi_p)}_{0}}\vk\cdot\left(
\vk_{2}-\frac{\ \vk_{2}\cdot\vec{\tau_p}}{ {(\vPi_p)}_{0}^2}\vec{\tau_p}
\right),
\end{equation}
\begin{equation}\label{gamma2}
\,^{\rm eik.}\gamma_{2p-1}^{bc}(\vk,\vk_{2})=-\delta_{2p-1}^{b}\delta_{2p}^{c}\ \frac{m a}{2k_{2}^{2}{(\vPi_p)}_{0}}\vk\cdot\left(
\vk_{2}-\frac{\ \vk_{2}\cdot\vec{\tau_p}}{ {(\vPi_p)}_{0}^2}\vec{\tau_p}
\right).
\end{equation}
This expression depends on each flow through its initial momentum $\vec{\tau}_p$, which vanishes in the standard non-relativistic equations. Besides, $(\vPi_p)_{0} \rightarrow - ma$ in the non-relativistic limit so that one recovers the expected formulae in this limit (see \cite{2010PhRvD..82h3520T,2013arXiv1311.2724B}).

In the following, we exploit the consequences of this approximation in order to evaluate the relevance of the coupling terms.
As a first step, we can notice from eqs. (\ref{fb:Xidef}), (\ref{gamma1}) and (\ref{gamma2}) that the two non-zero elements  coming from the flow labeled by $\vPi_{p}$ in the $\Xi$ matrix, $\Xi_{2p}^{\ \ 2p}$ and $\Xi_{2p-1}^{\ \ 2p-1}$,
are proportional to the velocity divergence of large-scale modes. Thus we can write
\begin{equation}\label{displacement}
\int_{\ze_{0}}^{\ze}\Xi_{a}^{\ b}(\ze',\vk)\dd\ze'=\ii\,\vk.\vd_{p}(\ze_{0},\ze)\ \delta_{a}^{\ b},
\end{equation}
where $a$ and $b$ are either $2p$ or $2p-1$ and where $d_{p}$ is the total displacement field induced by the large-scale modes in the fluid labeled by $\vPi_{p}$. It reads necessarily
\begin{equation}
\label{displacementp}
\vd_{p}(\ze,\ze_{0})=\ii\int_{\ze_{0}}^{\ze}\dd\ze'\int\dd^{3}\vq\frac{m a}{q^{2}{(\vPi_p)}_{0}} \left(\vq-\frac{\ \vq\cdot\vec{\tau_p}}{ {(\vPi_p)}_{0}^2}\vec{\tau_p}\right)\Psi_{2 p}(\ze',\vq).
\end{equation} 
Note that this displacement is superimposed on the zeroth order displacement field induced by the homogeneous momentum
of the flow. This background displacement field is given by (see eqs. (\ref{Fouriersub1}) and (\ref{Fouriersub2})) 
\begin{equation}\label{displacement0}
\vd_{p}^{(0)}=-\int_{\ze_{0}}^{\ze}\frac{\vec \tau_{p}}{\mH(\tau_{p})_{0}}.
\end{equation}
We will discuss this point in greater detail in the next section.

The impact of the eikonal correction introduced in the equation of motion depends on the way in which the various large-scale 
modes contribute to  the displacement fields $\vd_{p}$. The nature of the displacement is crucial in this context. In particular, one expects global displacement fields, which affect all species in a similar way, to induce a mere phase shift in the solution of the eikonal limit of (\ref{EikEoM}). Remarkably, such displacements have no impact on power spectra, provided that the fields at play are evaluated at the same time (see the end of this section). Using the standard language of cosmology, we call those particular displacements ``adiabatic displacements'' in the following. On the contrary, relative displacements between species can induce a damping in power spectra. It is simply due to the fact that species must evolve in phase (at least during a small period of time) for couplings between them to generate a significant growth of perturbations. This phenomenon has been highlighted for the first time in \cite{2010PhRvD..82h3520T} an reconsidered in \cite{2013PhRvD..87d3530B}. In this paper, we show that such considerations can be extended to the study of relativistic fluids. 

Knowing this, it is convenient to decompose displacement fields into

\begin{equation}\label{displacements}
d_{p}(\ze,\ze_{0})=d_{\adiab}(\ze,\ze_{0})+\delta d_{p}(\ze,\ze_{0}), 
\end{equation}
where $d_{\adiab}$ denotes naturally the adiabatic part. The other part, $\delta d_{p}$, represents relative motions between fluids. For non-relativistic species, it is known that the most growing mode is part of the adiabatic modes. Since neutrinos become non-relativistic at late time, one expects the most growing mode of each fluid of neutrinos to become also an adiabatic mode ultimately.

When considering only the adiabatic part of the displacement field, the solution of the eikonal equation of motion is easy to find. Its form is related to the extended Galilean invariance of the equations of motion, well established for non-relativistic species and uncovered for the system (\ref{Fouriersub1})-(\ref{Fouriersub2}) in \cite{2014arXiv1411.0428D}. Indeed, it has been shown that this system is invariant under the following transformations,
\begin{eqnarray}
\tilde x^{i}&=&x^{i}+d_{i}(z),\\
\tilde z&=&z,\\
\label{Gtransformationf}
\tilde\psi_{a}(\tilde z,\tilde \vx)&=&\psi_{a}(\eta,\vx),
\end{eqnarray}
where the last transformation can equivalently be written $\tilde\psi_{a}(\tilde\eta,\vk)=\exp(\ii \vk\cdot\vd(\eta))\psi_{a}(\eta,\vk)$.
It means that a homogeneous time-dependent displacement which disrupts the medium can be re-absorbed 
in a global phase shift of the Fourier transforms of the fields. Here, we are interested in making the large-scale adiabatic displacement fields play the role of the disturbers of the medium. 

More explicitly, it is known that the solution of the standard linear system can be fully described with the help of its Green function, $g_{a}^{\ b}(\ze,\ze_{0};\vk)$, defined in such a way that (see appendix \ref{Integral form} for details)
\begin{equation}
\Psi_{a}(\ze,\vk)=g_{a}^{\ b}(\ze,\ze_{0};\vk)\Psi_{b}(\ze_{0},\vk),
\end{equation}
with $\ze$ and $\ze_{0}$ two arbitrary times. Besides, when the displacement field is purely adiabatic, the Green functions  $\xi_{a}^{\ b}$ of the corrected linear system (i.e. the linear system in which an eikonal correction has been added) are related to those of the naked theory by
\begin{equation}
\xi_{a}^{\ b}(\ze,\ze_{0};\vk)= g_{a}^{\ b}(\ze,\ze_{0};\vk)\exp(\ii\vk.\vd_{\adiab}(\ze,\ze_{0})).
\end{equation}
A direct consequence of the symmetry property is that, after a transformation of the form (\ref{Gtransformationf}), ensemble averages of any product of fields $\psi_{a}(z_{a},\vk_{a})$ become proportional to \\ $\exp[\ii\sum_{a}\vk_{a}\cdot \vd_{\adiab}(z_{a})]$, which is unity when the fields are computed at equal time. Indeed, statistical homogeneity imposes $\sum_{a}\vk_{a}=0$.  This is the reason why equal time spectra or poly-spectra are not sensitive to the presence of adiabatic displacements.

Yet, we are interested in all contributions to the displacement field, including those that induce large motions between species.
As already mentioned, in early-time mixtures of baryons and cold dark matter, those relative displacements (i.e. those non-adiabatic contributions that 
develop in the nonlinear regime) are the leading  contributions to the nonlinear evolution of the power spectrum. To sketch the impact 
of massive neutrinos on the nonlinear growth rate, we evaluate in the next section the amplitude of the relative displacements involving them and 
we compute the corresponding power spectra.

\section{Relative displacements and power spectra: quantitative results}\label{power spectra}


As stressed in the previous section, in the eikonal approximation, the displacement fields that can cause a damping of the growth of structure are those that differ from one fluid to another. Hence, in this section, we compute on the one hand the power spectrum of the total displacement field of cold dark matter and, on the other hand, the power spectra of the relative displacement fields of relativistic flows (with respect to the a priori dominant cold dark matter component). Note that, in perturbation theory, such displacements are sums of terms of different orders (the zeroth order contribution $\vd^{(0)}_{p}$ being defined
in eq. (\ref{displacement0})),
\begin{equation}
\vd_{p}(\ze,\ze_{0})=\vd^{(0)}_{p}(\ze,\ze_{0})+\vd^{(1)}_{p}(\ze,\ze_{0})+\dots
\end{equation}
Note also that the damping due to non-homogeneous corrections can be significant only if the considered flow is non-relativistic (since in that case $\vd^{(0)}_{\vPi}(\ze,\ze_{0})$ is small) 
or if $\vk $ is orthogonal to the zeroth order contribution (since in that case $\vk.\vd^{(0)}_{\vPi}(\ze,\ze_{0})$ is small compared with the other contributions). Hence, in the following, we compute the expected values of $\vk.(\vd_{p}-\vd_{\cdm})$ as a function of the angle between the initial momentum $\vec{\tau}$ of the considered fluid  and $\vk$. Calculations are performed in the conformal Newtonian gauge,
\begin{equation}\label{metricConf}
\dd{s}^2=a^2\left(\eta\right)\left[-\left(1+2\psi\right)\dd{\eta}^2+\left(1-2\phi\right)\dd{x^i}\dd{x^{j}}\delta_{ij}\right].
\end{equation}
This choice will allow us to take advantage of the numerical work presented in \cite{2014JCAP...01..030D}.
First, let us define two transfer functions $D_{\cdm}$ and $D^{(\alpha)}_{p}$ as
\begin{eqnarray}
\int_{\ze_{0}}^{\ze}\dd\ze'\dfrac{ma}{{-(\tau_p)}_0}\ \left(\frac{\tau_p}{{(\tau_p)}_0}\right)^{\alpha}\,\theta_{\vPi_{p}}(\ze',\vq)&=&D^{(\alpha)}_{p}(\ze,\ze_{0},\vq)\psi_{\rm init}(\vq),
\\
\int_{\ze_{0}}^{\ze}\dd\ze'\ \theta_{\cdm}(\ze',\vq)&=&D_{\cdm}(\ze,\ze_{0},\vq)\psi_{\rm init}(\vq),
\end{eqnarray}
where $\psi_{\rm init}$ is the initial value of the potential $\psi$. 
The statistical properties of those quantities are entirely encoded in their initial power spectra $P_{\psi}(q)$, defined so that
\begin{equation}
\langle \psi_{\rm init}(\vq) \psi_{\rm init}(\vq')\rangle=(2\pi)^{3}\Dirac(\vq+\vq')\ P_{\psi}(q).
\end{equation}

Using the expressions of the transfer functions, the contribution of each mode to the displacement field of each relativistic flow reads
\begin{equation}
\vd_{p}(\ze,\ze_{0};\vq)=-\ii\left(\frac{\vq}{q^{2}} D^{(0)}_{{p}}(\ze,\ze_{0},\vq)-\frac{\vec{\tau}_p.\vq}{q^{2}(\tau_p)^2}\vec{\tau}_p D^{(2)}_{{p}}(\ze,\ze_{0},\vq)\right)\psi_{\rm init}.
\end{equation}
Besides, for the cold dark matter component, one simply has
\begin{equation}
\vd_{\cdm}(\ze,\ze_{0};\vq)=-\ii\frac{\vq}{q^{2}}\,D_{\cdm}(\ze,\ze_{0},\vq)\psi_{\rm init}(q).
\end{equation}

Furthermore, one can notice that the displacement field of a relativistic flow along an arbitrary direction $\vk$ depends on the angles between both $\vq$ and  $\vec\tau$  and $\vk$ and $\vec{\tau}$. After integration over the other angles, one finds for the variances of respectively  $\vk.\vd_{\cdm}$ and $\vk.(\vd_{p}-\vd_{\cdm})$,
\begin{equation}\label{cdm}
\langle (\vk.\vd_{\cdm})^{2}\rangle
=4\pi k^2\,\int \dd q \,P_{\psi}(q)\,\frac{1}{3}\ 
\vert D_{\cdm}(\vq)\vert^{2}
\end{equation}
and
\begin{eqnarray}\label{relDispPower}
&&\langle (\vk.(\vd_{p}-\vd_{\cdm}))^{2}\rangle
= 2\pi k^2\times \\&&\int \dd q \,P_{\psi}(q)\int_{-1}^{1}\dd \mu \,\nonumber 
\left[
\frac{1}{2}(1-\mu^{2}_{k})(1-\mu^{2})\vert D^{(0)}_{p}(\vq)-D_{\cdm}(\vq)-D^{(2)}_{p}(\vq)\vert^{2}+\mu^{2}\mu^{2}_{k}
\vert D^{(2)}_{p}(\vq)\vert^{2}
\right],
\end{eqnarray}
where $\mu_k$ is the Cosine of the angle between the initial momentum of the flow $\vec\tau_{p}$ and $\vk$ and where  
an integration is made over $\mu$, Cosine of the angle between $\vec\tau_{p}$ and $\vq$.
In eq. (\ref{relDispPower}), one can notice that the dependence of the r.m.s. with respect to $\mu_{k}$ is such that it does not vanish either for an initial
momentum $\vec\tau_{p}$ orthogonal to $\vk$ (i.e. when $\mu_{k}=0$) or for an initial flow momentum along $\vk$ (i.e. when $\mu_{k}=1$). 

%
%
%

\begin{figure}[!h]
\centering
\includegraphics[width=7cm]{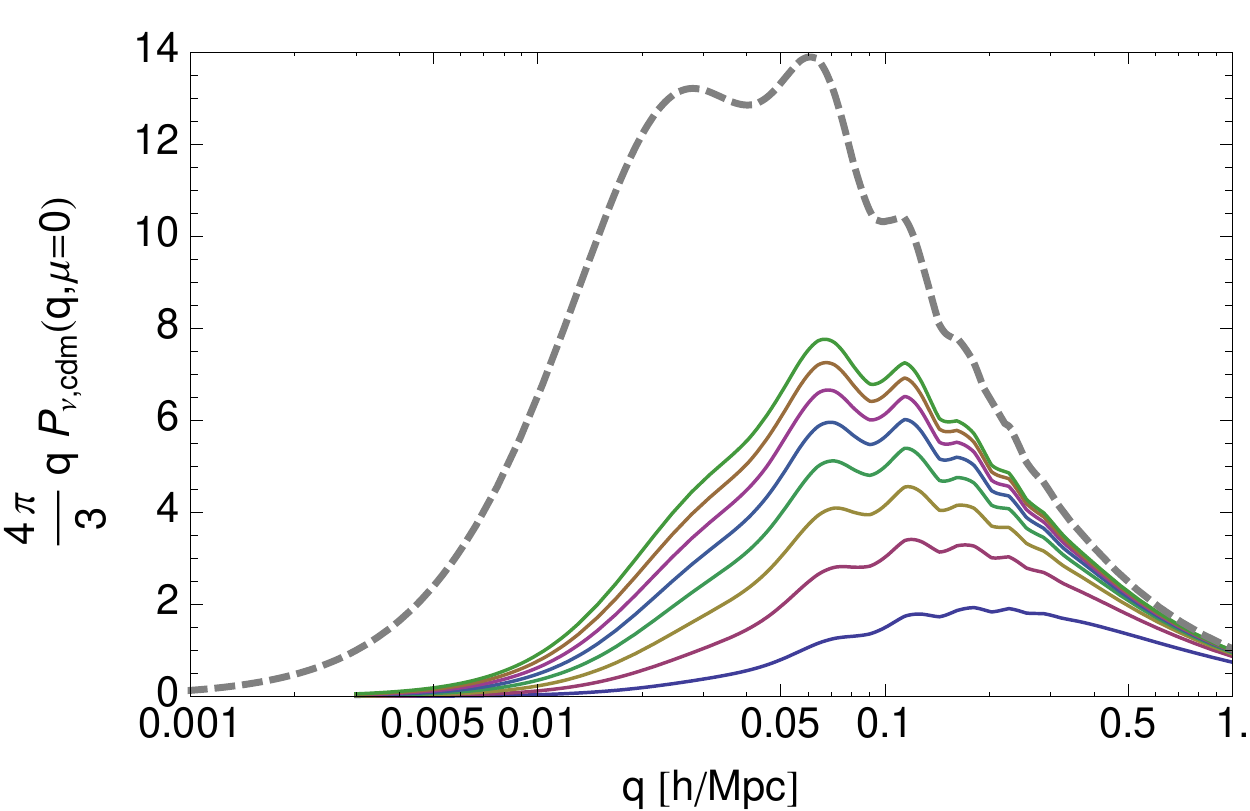} 
\includegraphics[width=7cm]{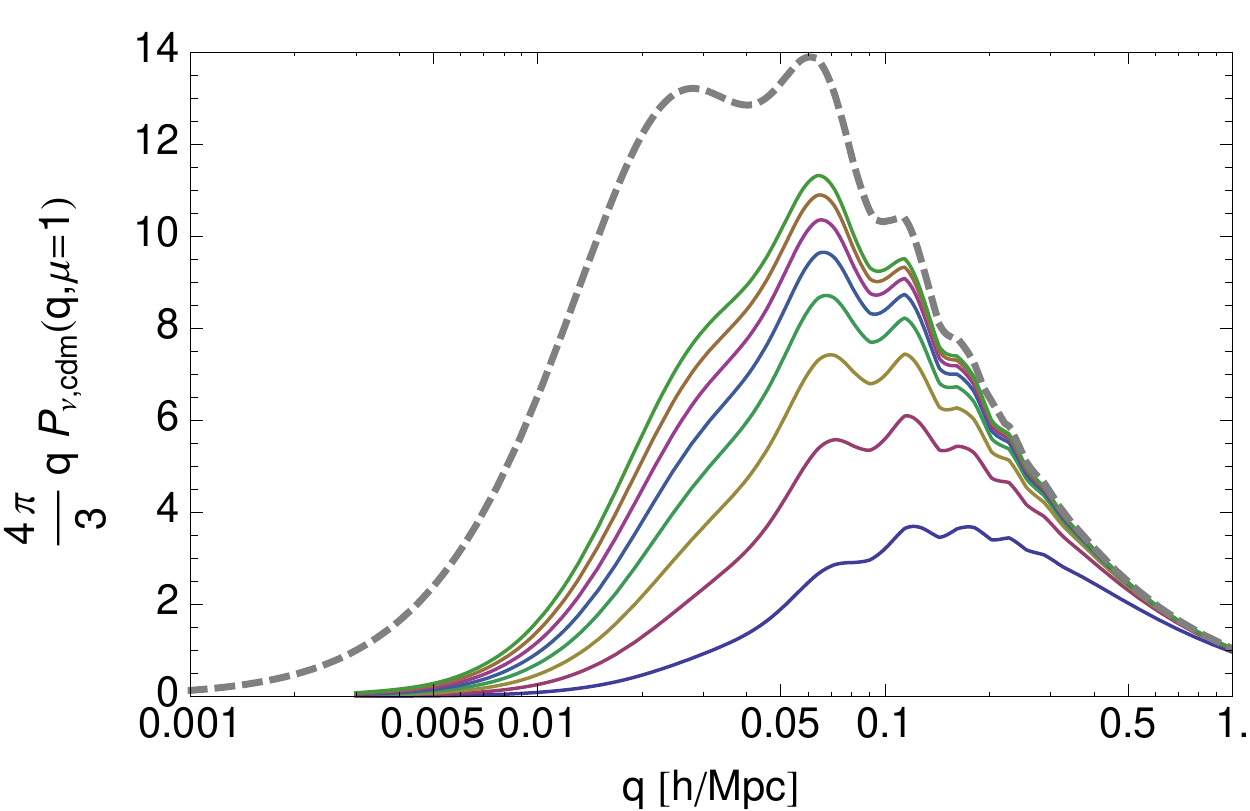} 
\caption{Power spectrum of the relative displacement as a function of the mode $q$ for different neutrino flows. The quantities $P_{\rm \nu, cdm}$ are defined so that the right hand sides of eqs. (\ref{cdm}) and (\ref{relDispPower}) read $4 \pi k^2/3 \int{\dd q P_{\rm \nu, cdm}}(q)$.
The values of $\vPi$ range from 2.25 $k_{B}T_{0}$ (bottom line) to 18 $k_{B} T_{0}$ (top line).  On the left panel $\mu_{k}$ is set to 0, on the right panel $\mu_{k}$ is set to 1 and the neutrino mass is set to $0.3 \ $eV. The gray dashed line represents the power spectrum of the cold dark matter displacement.}
\label{dPlot2}
\end{figure}

On Figure \ref{dPlot2}, we present the per mode contribution to the right hand side of eq. (\ref{relDispPower})  in the particular case of neutrino fluids for $\mu_{k}=0$ on the left panel and $\mu_{k}=1$
on the right panel. 
The results have been computed assuming a single species of
neutrinos whose mass is $m_{\nu}=0.3\ $eV and using the cosmological parameters derived from the Five-Year Wilkinson Microwave Anisotropy Probe (WMAP 5) observations. Besides, the values of the  initial power spectra have been obtained under the assumption that, in the cosmological model we adopt, metric fluctuations are initially adiabatic and characterized by the scalar spectral index $n_{s}\approx 0.96$.  We can see that the resulting neutrino power spectra are comparable in amplitude to the cold dark matter one
(represented by a thick dashed line).

Denoting $\sigma_{d_{\cdm}}=1/k \langle (\vk.\vd_{\cdm})^{2}\rangle^{1/2}$ and $\sigma_{d_{\vPi}}=1/k\langle (\vk.(\vd_{p}-\vd_{\cdm}))^{2}\rangle^{1/2}$
we find that  $\sigma_{d_{\vPi}}$ is of the order of 2 to about 5$h^{-1}$Mpc for the cosmological modes we used for which $\sigma_{d_{\cdm}}\approx 6h^{-1}$Mpc,. This result  depends of course on the flow considered and is actually of the order of the cold dark matter value.
What does it mean? Similarly to what happens in mixtures of  baryons and cold dark matter, one expects the perturbation growth
to be damped for wave numbers larger than or comparable to $1/\sigma_{d_{\vPi}}$, for which $\vk\cdot\vd^{(1)}_{\vPi}$ is expected to be finite.  Besides, such a non-adiabatic damping is potentially larger than 
the damping due to the homogeneous displacements $\vd_{p}^{(0)}$ of each flow when $\mu_{k}$ is close to zero.  A precise determination of the amplitude of these effects would require a full analysis of the nonlinear evolution of the system. We leave this for a future study.

\section{Conclusion}

Describing neutrinos as a collection of single-stream fluids is an efficient strategy to infer their impact on the growth of the cosmic structure. Using this approach, one indeed gets a complete set of equations of motion that incorporate all  the nonlinear effects of relativistic or non-relativistic particles. In the subhorizon limit, the system takes the form of
eq. (\ref{FullEoM}), which can be easily handled by a formalism originally developed to depict non-relativistic species.

In this paper, we evaluated the amplitudes of the nonlinear couplings and determined for each flow the scales at which they are expected to impact significantly on the structure growth. For that purpose, we implemented the eikonal approximation into the general equation of motion. We concluded that the impact of large-scale modes on an arbitrary mode are entirely driven by large-scale displacement fields whose expressions are given in eq. (\ref{displacementp}). The comparison between the displacement field associated with each flow of neutrinos and the one associated with cold dark matter makes easy the comparison between the power spectra of the relative displacements between neutrinos and cold dark matter and the power spectrum of the displacement of cold dark matter alone. We found as a preliminary result that couplings involving massive neutrinos (with a $0.3$ eV mass) are expected to induce a damping of the perturbation growth in neutrino flows for wave numbers larger than (or of the order of) about $0.2$ to $0.5\  h/$Mpc. A detailed quantitative analysis of the consequences of this phenomenon is yet to be done but those findings confirm the significance of nonlinear couplings in the dynamical evolution of neutrino fluids. This sets the stage for further numerical
studies  beyond the linear regime.

\medskip

\textbf{Acknowledgements}:  This work is partially supported by the grant ANR-12-BS05-0002 of the French Agence Nationale de la Recherche.

\newpage
\appendix
\section{Integral form of the overall equation of motion}
\label{Integral form}

For a finite number of flows, the overall equation of motion 
(\ref{FullEoM}) can formally be written in an integral form. It requires the use of the associated Green
operator, $g_{a}^{\ b}(\ze,\ze_{0};\vk)$,
of the linear system. This operator satisfies
\begin{equation}
\Psi_{a}(\ze,\vk)=g_{a}^{\ b}(\ze,\ze_{0};\vk)\Psi_{b}(\ze_{0},\vk),
\end{equation}
with $\ze$ and $\ze_{0}$ two arbitrary times. 
It is besides solution of the differential equation
\begin{equation}
\frac{\partial}{\partial\ze}g_{a}^{\ b}(\ze,\ze_{0};\vk)+\Omega_{a}^{\ c}(\ze;\vk)\,g_{c}^{\ b}(\ze,\ze_{0};\vk)=0
\end{equation}
with the condition
\begin{equation}
g_{a}^{\ b}(\ze_{0},\ze_{0};\vk)=\delta_{a}^{\ b},
\label{gabbound}
\end{equation}
$\delta_{a}^{\ b}$ being the identity matrix. Formally,  the Green
function is  the ensemble of all the independent 
linear solutions of the system\footnote{A priori the number of solutions is equal to twice the number of flows considered.}. Denoting $u_{a}^{(\alpha)}(\ze,\vk)$ these solutions, $g_{a}^{b}$ reads
\begin{equation}
g_{a}^{b}(\ze,\ze_{0},\vk)=\sum_{\alpha}u_{a}^{(\alpha)}(\ze,\vk)c_{(\alpha)}^{b}(\ze_{0},\vk),
\end{equation}
where the variables $c_{(\alpha)}^{b}(\ze_{0},\vk)$ are set so that (\ref{gabbound}) is satisfied.

Studying in detail the Green operator of such a system 
is beyond the scope of this appendix. Suffice to note here that, unlike the case of a single pressureless
flow,  the Green operator generally depends on the wave mode $k$. This dependence is expected to gradually decay over time and to disappear at very late time, when all the flows have become non-relativistic. At this stage, the situation is then identical to the one of  a collection of cold dark matter fluids.

As for the standard system of non-relativistic particles, the knowledge of the Green operator of the equation of motion allows to write a formal
solution (see \cite{1998MNRAS.299.1097S,2001NYASA.927...13S,2006PhRvD..73f3519C}), which is given by
\begin{eqnarray}
\Psi_a(\vk,\ze) &=& g_{a}^{\ b}(\vk,\ze,\ze_{0}) \ \Psi_b(\vk,\ze_{0}) + \nonumber\\
&+&\int_{\ze_{0}}^{\ze}  {\dd \ze'} \ g_{a}^{\ b}(\vk,\ze,\ze') \ 
\gamma_{b}^{\ cd}(\vk_1,\vk_2) \Psi_c(\vk_1,\ze') \Psi_d(\vk_2,\ze'),
\label{eomi}
\end{eqnarray}
with $\Psi_a(\vk,\ze_0)$ the initial conditions. Many of the approaches developed in order to improve upon standard Perturbation Theory rely on an accurate description of the Green functions beyond the linear regime. This is the purpose for instance of RPT and RegPT methods (\cite{2006PhRvD..73f3519C,2014PhRvD..89b3502B,2012PhRvD..86j3528T}).

\bibliographystyle{JHEP}
\bibliography{HD_FB,LesHouches}
\end{document}